# The Equilibrium Climate Response to Sulfur Dioxide and Carbonaceous Aerosol Emissions from East and Southeast Asia

**Benjamin S. Grandey[1], Lik Khian Yeo[1,2], Hsiang-He Lee[1], and Chien Wang[3,1]**

[1]Center for Environmental Sensing and Modeling, Singapore-MIT Alliance for Research and Technology, Singapore, Singapore.

[2]Department of Civil and Environmental Engineering, National University of Singapore, Singapore, Singapore.

[3]Center for Global Change Science, Massachusetts Institute of Technology, Cambridge, Massachusetts, USA.

Corresponding authors: Benjamin S.Grandey (benjamin@smart.mit.edu) and Chien Wang (wangc@mit.edu)

**Key Points:**

- Anthropogenic aerosol emissions from East and Southeast Asia exert a global mean net radiative effect of -0.49±0.04 W m$^{-2}$

- Strong suppression of precipitation occurs over East and Southeast Asia

- We find no clear evidence of remote effects on precipitation over Australia and West Africa

**Abstract**

We investigate the equilibrium climate response to East and Southeast Asian emissions of carbonaceous aerosols and sulfur dioxide, a precursor of sulfate aerosol. Using the Community Earth System Model, with the Community Atmosphere Model version 5.3, we find that anthropogenic aerosol emissions from East and Southeast Asia exert a global mean net radiative effect of -0.49±0.04 W m$^{-2}$. Approximately half of this cooling effect can be attributed to anthropogenic sulfur dioxide emissions. The aerosol emissions drive widespread cooling across the Northern Hemisphere. Strong suppression of precipitation occurs over East and Southeast Asia, indicating that anthropogenic aerosol emissions may impact water resources locally. However, in contrast to previous research, we find no clear evidence of remote effects on precipitation over Australia and West Africa. We recommend further investigation of possible remote effects.

**Plain Language Summary**

Aerosols – particles suspended in the atmosphere – influence clouds and the climate system. Using a state-of-the-art global climate model, we investigate the climate impacts of aerosol emissions from human activity in East and Southeast Asia. We find that the aerosol emissions lead to widespread cooling across the Northern Hemisphere. The aerosol emissions also suppress rainfall over East and Southeast Asia, potentially impacting water resources.

**1 Introduction**

Anthropogenic emissions from East and Southeast Asia are responsible for 21% of global sulfur dioxide emissions, 13% of global organic carbon aerosol emissions, and 26% of global black carbon aerosol emissions (Fig. 1). These emissions likely influence the climate system by interacting directly with radiation (Haywood & Boucher, 2000) and by interacting with clouds (Fan et al., 2016; Rosenfeld et al., 2014).

In contrast to well-mixed greenhouse gases, the radiative effects of aerosols are regionally heterogeneous. The regional heterogeneity of the aerosol radiative effects likely influences surface temperature gradients and the large-scale distribution of precipitation (Wang, 2015).

Monsoon systems may be particularly sensitive to the heterogeneity of the aerosol radiative effects. For example, several studies have suggested that anthropogenic aerosols influence the East Asian monsoon: the distribution of precipitation over China may be influenced both by absorbing aerosols, such as black carbon (Gu et al., 2006; Jiang et al., 2013; Lau et al., 2006; Menon, 2002), and by sulfate aerosol, via indirect effects on clouds (Guo et al., 2013; Jiang et al., 2013; Liu et al., 2011).

It is likely that the South Asian monsoon is sensitive to absorbing aerosols such as black carbon (Chung et al., 2002; Lau et al., 2006; Lee et al., 2013; Lee & Wang, 2015; Meehl et al., 2008; Ramanathan et al., 2005; Wang et al., 2009); and the South Asian monsoon may be even more sensitive to the indirect effects of sulfate aerosol (Bollasina et al., 2011). Alongside local emissions sources, remote emissions sources may contribute to the influence of aerosols on the South Asian monsoon (Bollasina et al., 2014; Cowan & Cai, 2011; Ganguly et al., 2012).

East and Southeast Asian emissions may influence precipitation remotely in other regions by perturbing large-scale circulations (Bartlett et al., 2018). Previous studies have suggested that

___

Northern Hemisphere aerosol emissions remotely influence the Australian monsoon (Grandey et al., 2016; Rotstayn et al., 2012) and the West African monsoon (Bartlett et al., 2018; Booth et al., 2012; Dong et al., 2014; Grandey et al., 2016).

In this letter, we explore the climate response to anthropogenic aerosol emissions from East and Southeast Asia. Alongside organic carbon aerosol and black carbon aerosol, we also consider sulfur dioxide, a precursor of sulfate aerosol.

Based on the exploratory research of Grandey et al. (2016) – who investigated the transient climate response to aerosol emissions from a larger Asia region – we hypothesise that the anthropogenic aerosol emissions from East and Southeast Asia will (1) drive widespread cooling of the Northern Hemisphere, leading to interhemispheric asymmetry in the surface temperature response, (2) suppress precipitation in the Northern Hemisphere tropics and weakly enhance precipitation in the Southern Hemisphere tropics, (3) suppress East Asian monsoon precipitation, (4) suppress South Asian monsoon precipitation, especially over southern India, (5) enhance Australian monsoon precipitation, and (6) suppress West African monsoon precipitation over the Sahel. These hypotheses are discussed and tested in Section 4.

**2. Methods**

2.1 Scenarios

The three scenarios differ only in their aerosol emissions (Fig. 1):

1. *Ref*, the reference scenario, uses year-2000 emissions (Supplement of Lamarque et al., 2010; Liu et al., 2012).

2. *Exp1*, the first experimental scenario, has no anthropogenic emissions of sulfur dioxide (including primary sulfate) from East and Southeast Asia: sulfur dioxide emissions from all anthropogenic sectors, including shipping, are removed within the regional bounds of 94–161°E, 10°S–65°N. Comparison of *Exp1* with *Ref* reveals the influence of anthropogenic sulfur dioxide emissions from East and Southeast Asia, assuming all other emissions remain at year-2000 levels.

3. *Exp2*, the second experimental scenario, has no anthropogenic emissions of aerosol (sulfur dioxide, organic carbon, black carbon) from East and Southeast Asia. Comparison of *Exp2* with *Ref* reveals the influence of anthropogenic aerosol emissions from East and Southeast Asia.

2.2 Model Configuration

Simulations are performed using the Community Earth System Model version 1.2.2 (CESM1.2.2), with the Community Atmosphere Model version 5.3 (CAM5.3) and an aerosol module with three log-normal modes (MAM3) (Liu et al., 2012). CAM5.3 is run at a horizontal resolution of 1.9°×2.5° with 30 levels in the vertical direction. Greenhouse gas concentrations are prescribed using year-2000 climatological values.

For each emissions scenario, two simulations are performed: a simulation using prescribed sea surface temperatures (SSTs); and a simulation using a coupled atmosphere–ocean configuration.

The prescribed-SST simulations facilitate diagnosis of aerosol radiative effects (Ghan, 2013; Section 2.4 of Grandey et al., 2018) . The configuration follows the "F_2000_CAM5" component set. Each prescribed-SST simulation is run for 32 years: the first two years are excluded from the analysis; the final 30 years are analyzed.

The coupled atmosphere–ocean simulations facilitate investigation of the equilibrium climate response to changes in the aerosol emissions. The configuration follows the "B_2000_CAM5_CN" component set. The three-dimensional ocean model uses a displaced pole grid with a resolution of approximately 1°×1°. Each coupled atmosphere–ocean simulation is run for 100 years: the first 40 years, during which the simulation moves towards a near-equilibrium state, are excluded from the analysis; the final 60 years are analyzed.

## 3. Results

### 3.1 Radiative Effects

The anthropogenic sulfur dioxide emissions from East and Southeast Asia exert a net cooling effect on the climate system, contributing -0.24±0.04 W m$^{-2}$ to the net effective radiative forcing (Fig. 2a). The net radiative effect is dominated by the shortwave cloud radiative effect (-0.44±0.04 W m$^{-2}$; Fig. S1a) and the longwave cloud radiative effect (+0.26±0.02 W m$^{-2}$; Fig. S2a). The direct radiative effect (-0.03±0.01 W m$^{-2}$; Fig. S3a) and the surface albedo radiative effect (-0.04±0.02 W m$^{-2}$; Fig. S4a) are much smaller.

When carbonaceous aerosol emissions are also modified, the anthropogenic aerosol emissions exert an even stronger net radiative effect of -0.49±0.04 W m$^{-2}$, twice as strong as when only the sulfur dioxide emissions are modified. The shortwave cloud radiative effect becomes stronger (-0.66±0.03 W m$^{-2}$): in MAM3, organic carbon aerosol is internally-mixed with other species of high hygrocopicity and therefore contributes efficiently to cloud condensation nuclei availability (Grandey et al., 2018). Counterintuitively, the longwave cloud radiative effect becomes weaker (+0.18±0.02 W m$^{-2}$): this is unexpected, because neither organic carbon aerosol nor black carbon aerosol acts as ice nuclei in CAM5.3 (Gettelman et al., 2010). The direct radiative effect becomes positive (+0.03±0.01 W m$^{-2}$), due to absorption by black carbon aerosol.

The longwave cloud radiative effect (Fig. S2b) of the anthropogenic aerosol emissions is strongest over Southeast Asia, the South China Sea, the Bay of Bengal, and the eastern Indian Ocean – in the tropics. In contrast, the shortwave cloud radiative effect (Fig. S1b) and the net radiative effect (Fig. 2b) are strongest over East Asia and the western North Pacific Ocean – in the Northern Hemisphere subtropics and midlatitudes. The net radiative effect exhibits strong interhemispheric asymmetry.

### 3.2 Surface Temperature

The interhemispheric asymmetry in the net radiative effect drives interhemispheric asymmetry in the surface temperature response (Fig. 3). The East and Southeast Asian aerosol emissions drive widespread cooling of the Northern Hemisphere, especially in the mid-latitudes and the Arctic. The aerosol emissions also cool parts of the Southern Hemisphere, but the cooling is generally not as strong as in the Northern Hemisphere.

When both carbonaceous and sulfur emissions are modified, the global mean cooling (-0.19±0.03 °C) is approximately twice as strong as when only sulfur emissions are modified (-0.10±0.03 °C). As mentioned above, the net radiative effects for the two experimental cases also differ by a factor of two. Hence the relationship between the net radiative effect and the global mean temperature response is linear: both experimental cases produce a sensitivity of -0.4 °C (W m$^{-2}$)$^{-1}$.

3.3 Precipitation

Precipitation is sensitive to changes in surface temperature gradients (Chiang & Friedman, 2012; Wang, 2015). The interhemispheric asymmetry in the surface temperature response drives interhemispheric asymmetry in the precipitation response: in particular, the East and Southeast Asian aerosol emissions suppress precipitation in the Northern Hemisphere (Fig. 4a). Precipitation is strongly suppressed over East and Southeast Asia; precipitation is enhanced over the western Indian Ocean and over parts of the eastern Pacific Ocean, but statistical significance is generally absent (Fig. 4b). We further discuss the regional precipitation response in the Section 4 below.

**4 Discussion and Recommendations**

In Section 1 we stated six hypotheses regarding the influence of anthropogenic aerosol emissions from East and Southeast Asia, based on the conclusions of Grandey et al. (2016). We now discuss and test these hypotheses.

First, we hypothesized that the emissions would drive widespread cooling of the Northern Hemisphere, leading to interhemispheric asymmetry in the surface temperature response. Our results support this hypothesis: the interhemispheric asymmetry in the net radiative effect (Fig. 2) drives interhemispheric asymmetry in the surface temperature response (Fig. 3).

Second, we hypothesized that the emissions would (i) suppress precipitation in the Northern Hemisphere tropics and (ii) weakly enhance precipitation in the Southern Hemisphere tropics. Such a precipitation response would be consistent with a southward shift of the intertropical convergence zone in response to the cooling of the Northern Hemisphere (Chiang & Friedman, 2012). Our results support the first part of this hypothesis: precipitation is suppressed in the Northern Hemisphere tropics (Fig 4a). However, due to the large standard errors, our results do not provide conclusive evidence that precipitation is enhanced in the Southern Hemisphere tropics.

Third, we hypothesized that the aerosol emissions would suppress East Asian monsoon precipitation. Our results support this hypothesis: annual precipitation is suppressed over East Asia (Fig. 4b); the suppression occurs throughout the year, especially during the summer monsoon months (Fig. S5b). Surface cooling likely contributes to the suppression of precipitation (Jiang et al., 2013). In agreement with Grandey et al. (2016), the suppression of precipitation is also associated with anomalous downward motion (often indicating suppressed ascent) in the mid-troposphere between 25–35°N, although statistical significance is absent (Fig. 6b); anomalous upward motion (generally indicating suppressed descent) occurs between 35–45°N.

Fourth, we hypothesized that the aerosol emissions would suppress South Asian monsoon precipitation, especially over southern India. Our results provide only weak support for this hypothesis: annual precipitation is suppressed over southern India, but significance is absent (Fig. 4b). The suppression of precipitation over southern South Asia only occurs during January–July, with suppression over central South Asia during August, and enhancement over much of South Asia during September–December (Fig. S7b). (The precipitation response is correlated with the mid-tropospheric motion anomalies, shown in Fig. S8b.) Grandey et al. (2016) – who modified the aerosol emissions over South Asia alongside other parts of Asia – reported much stronger suppression of precipitation over southern South Asia: this discrepancy between the two studies suggests that local South Asian emissions contribute to the suppression reported by Grandey et al. (2016). In particular, local emissions of black carbon aerosol may impact the South Asian monsoon (Chung et al., 2002; Lau et al., 2006; Lee et al., 2013; Lee & Wang, 2015; Meehl et al., 2008; Ramanathan et al., 2005; Wang et al., 2009).

Fifth, we hypothesized that the aerosol emissions would enhance Australian monsoon precipitation. Our results do not support this hypothesis: although weak enhancement of annual precipitation occurs over western Australia (Fig. 4b), there is no clear enhancement of the austral monsoon precipitation or circulation over northern Australia (Figs. S9b, S10b).

Sixth, we hypothesized that the aerosol emissions would suppress West African monsoon precipitation over the Sahel. Again, our results do not support this hypothesis: annual precipitation is not suppressed over the Sahel (Fig. 4b); there is no suppression during the summer monsoon months, except during September (Fig. S11b). However, there are two similarities with the results of Grandey et al. (2016) in the vicinity of West Africa: the aerosols appear to drive a slight southward shift of the intertropical convergence zone (Fig. S11b) and a slight weakening of the West African westerly jet (Pu & Cook, 2012) during September and October (Fig. S12b), although significance is absent.

To summarize, we find interhemispheric asymmetry in the radiative effects, surface temperature response, and precipitation response, in agreement with the results of Grandey et al. (2016). We also find local suppression of precipitation over East Asia, again in agreement with Grandey et al. (2016). However, diverging from Grandey et al. (2016), we do not find clear evidence of remote effects, especially over Australia and the Sahel. Considering the methodological differences between the two studies (Table S1), we identify four possible reasons for this discrepancy between the studies: first, South Asian emissions may play an important role in remote effects outside Asia; second, non-linear interactions between aerosol emissions from Asia, aerosol emissions outside Asia, and greenhouse gas forcing may be partially responsible for some of the results of Grandey et al. (2016); third, our signal-to-noise ratio may be too weak to distinguish the remote effects identified by Grandey et al. (2016); and fourth, it is possible that some of the results of Grandey et al. (2016) are spurious.

We recommend further investigation of proposed remote effects. Exploratory research – such as that presented by Grandey et al. (2016) – helps us to formulate hypotheses, but the risk of spurious results should be acknowledged. Exploratory research should be followed by research that tests clearly formulated hypotheses, as we have attempted to do in this letter. Hypotheses should ultimately be tested using a range of different climate models. Methods that increase statistical rigor – such as the method advocated by Wilks (2016) – should also be implemented. Confirmation of hypotheses is valuable; negation of hypotheses is even more valuable, revealing the need for further research.

## 5. Summary and Conclusions

Our results suggest that anthropogenic aerosol emissions from East and Southeast Asia exert a net radiative effect of -0.49±0.04 W m$^{-2}$ on the climate system, largely due to indirect effects via clouds. Sulfur dioxide emissions are responsible for approximately half of the net radiative effect.

Although the radiative effects are concentrated in the vicinity of the source region, the sea-surface temperature response facilitates widespread cooling of the Northern Hemisphere. The cooling of the Northern Hemisphere drives suppression of rainfall in the Northern Hemisphere tropics.

Strong suppression of rainfall occurs over the source region of East and Southeast Asia. However, we find no clear evidence of remote effects on the Australian monsoon and the West African monsoon, although such remote effects cannot be ruled out. As discussed in the final paragraph of Section 4, we recommend further investigation of possible remote effects.

We conclude that anthropogenic aerosol emissions may influence rainfall across East and Southeast Asia. The potential impact on water resources in Southeast Asia is the focus of ongoing work.

**Code and Data Availability**

CESM 1.2.2 is available via http://www.cesm.ucar.edu/models/cesm1.2/. Model namelist files, configuration scripts, and analysis code are available via https://github.com/grandey/p17d-sulphur-eas-eqm, archived at https://doi.org/10.5281/zenodo.1211416. The input data and the model output data analysed in this paper are archived at https://doi.org/10.6084/m9.figshare.6072887.

**Author Contributions**

BSG and LKY designed the experiment, with contributions from HHL and CW. BSG configured the simulations and analyzed the results. BSG wrote the manuscript, with contributions from LKY, HHL, and CW.

**Acknowledgments**

This research is supported by the National Research Foundation of Singapore under its Campus for Research Excellence and Technological Enterprise programme. The Center for Environmental Sensing and Modeling is an interdisciplinary research group of the Singapore-MIT Alliance for Research and Technology. This research is also supported by the U.S. National Science Foundation (AGS-1339264) and the U.S. Department of Energy, Office of Science (DE-FG02-94ER61937). The CESM project is supported by the National Science Foundation and the Office of Science (BER) of the U.S. Department of Energy. We acknowledge high-performance computing support from Cheyenne (doi:10.5065/D6RX99HX) provided by NCAR's Computational and Information Systems Laboratory, sponsored by the National Science Foundation.

Preprint submitted to *arXiv*, 2018-04-20term climate? A case study for the Asian monsoon. *Climate Dynamics*, *50*(5–6), 1863–1880. http://doi.org/10.1007/s00382-017-3726-6

Benjamini, Y., & Hochberg, Y. (1995). Controlling the False Discovery Rate: A Practical and Powerful Approach to Multiple Testing. *J. R. Statist. Soc. B*, *57*(1), 289–300.

Bollasina, M. A., Ming, Y., & Ramaswamy, V. (2011). Anthropogenic Aerosols and the Weakening of the South Asian Summer Monsoon. *Science.*, *334*(6055), 502–505. http://doi.org/10.1126/science.1204994

Bollasina, M. A., Ming, Y., Ramaswamy, V., Schwarzkopf, M. D., & Naik, V. (2014). Contribution of local and remote anthropogenic aerosols to the twentieth century weakening of the South Asian Monsoon. *Geophysical Research Letters*, *41*(2), 680–687. http://doi.org/10.1002/2013GL058183

Booth, B. B. B., Dunstone, N. J., Halloran, P. R., Andrews, T., & Bellouin, N. (2012). Aerosols implicated as a prime driver of twentieth-century North Atlantic climate variability. *Nature*, *484*(7393), 228–232. http://doi.org/10.1038/nature10946

Chiang, J. C. H., & Friedman, A. R. (2012). Extratropical Cooling, Interhemispheric Thermal Gradients, and Tropical Climate Change. *Annual Review of Earth and Planetary Sciences*, *40*(1), 383–412. http://doi.org/10.1146/annurev-earth-042711-105545

Chung, C. E., Ramanathan, V., & Kiehl, J. T. (2002). Effects of the South Asian Absorbing Haze on the Northeast Monsoon and Surface–Air Heat Exchange. *Journal of Climate*, *15*(17), 2462–2476. http://doi.org/10.1175/1520-0442(2002)015<2462:EOTSAA>2.0.CO;2

Cowan, T., & Cai, W. (2011). The impact of Asian and non-Asian anthropogenic aerosols on 20th century Asian summer monsoon. *Geophysical Research Letters*, *38*(11). http://doi.org/10.1029/2011GL047268

Dong, B., Sutton, R. T., Highwood, E., & Wilcox, L. (2014). The Impacts of European and Asian Anthropogenic Sulfur Dioxide Emissions on Sahel Rainfall. *Journal of Climate*, *27*(18), 7000–7017. http://doi.org/10.1175/JCLI-D-13-00769.1

Fan, J., Wang, Y., Rosenfeld, D., & Liu, X. (2016). Review of Aerosol–Cloud Interactions: Mechanisms, Significance, and Challenges. *Journal of the Atmospheric Sciences*, *73*(11), 4221–4252. http://doi.org/10.1175/JAS-D-16-0037.1

Ganguly, D., Rasch, P. J., Wang, H., & Yoon, J.-H. (2012). Climate response of the South Asian monsoon system to anthropogenic aerosols. *Journal of Geophysical Research: Atmospheres*, *117*(D13). http://doi.org/10.1029/2012JD017508

Gettelman, A., Liu, X., Ghan, S. J., Morrison, H., Park, S., Conley, A. J., … Li, J.-L. F. (2010). Global simulations of ice nucleation and ice supersaturation with an improved cloud scheme in the Community Atmosphere Model. *Journal of Geophysical Research*, *115*(D18), D18216. http://doi.org/10.1029/2009JD013797

Ghan, S. J. (2013). Technical Note: Estimating aerosol effects on cloud radiative forcing. *Atmospheric Chemistry and Physics*, *13*(19), 9971–9974. http://doi.org/10.5194/acp-13-9971-2013

Grandey, B. S., Cheng, H., & Wang, C. (2016). Transient Climate Impacts for Scenarios of Aerosol Emissions from Asia: A Story of Coal versus Gas. *Journal of Climate*, *29*(8),8

## Figures

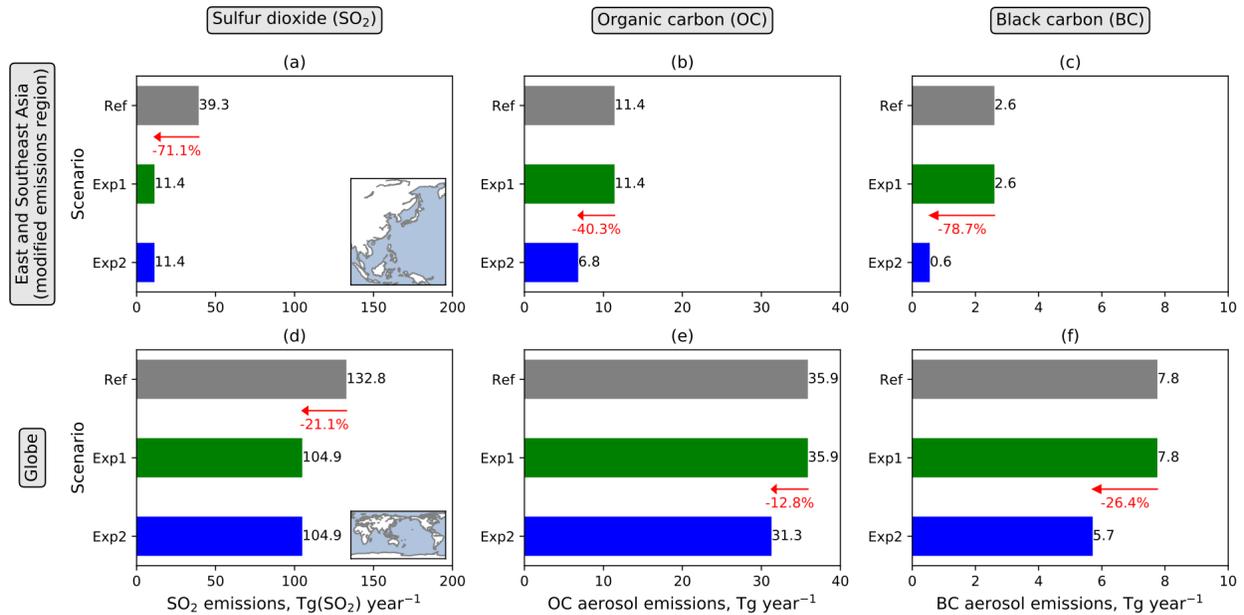

**Figure 1.** Annual emissions of sulfur dioxide (including primary sulfate), organic carbon aerosol, and black carbon aerosol for the three scenarios: *Ref* uses year-2000 aerosol emissions; *Exp1* has no anthropogenic sulfur emissions from East and Southeast Asia (94–161°E, 10°S–65°N); *Exp2* has no anthropogenic aerosol emissions from East and Southeast Asia. Red arrows and text indicate modifications to the emissions in the experimental scenarios. (a)–(c) Emissions from East and Southeast Asia; (d)–(f) global emissions. The emissions include both anthropogenic and natural sources. 2.5% of the sulfur dioxide is emitted as primary sulfate.

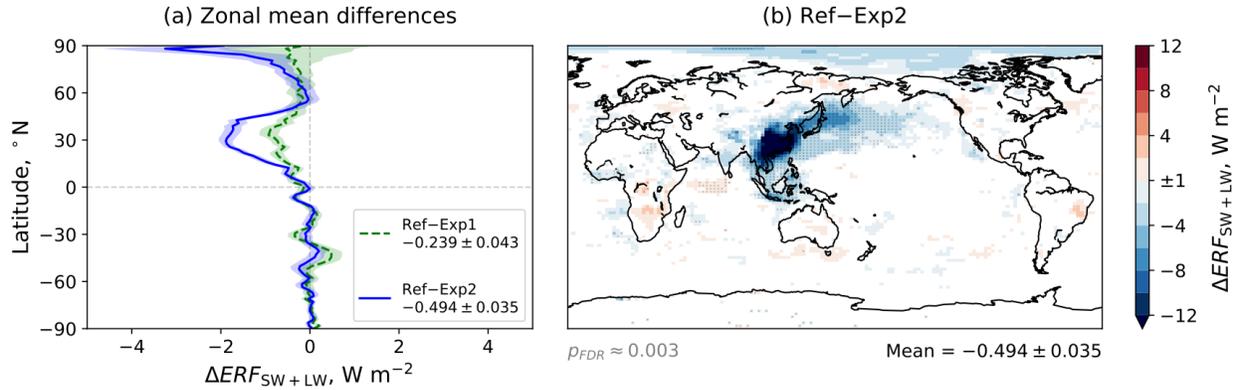

**Figure 2.** Differences in annual mean aerosol effective radiative forcing ($ERF_{SW+LW}$), diagnosed using the prescribed-SST simulations. (a) Zonal mean differences in $ERF_{SW+LW}$ for the experimental scenarios compared to the reference scenario (*Ref*). Shading indicates combined standard errors, calculated using the annual zonal means for each simulation year. The legend contains global area-weighted mean differences and associated combined standard errors, calculated using the annual global means for each simulation year. (b) Map of *Ref*–*Exp2* differences in $ERF_{SW+LW}$. White indicates differences with a magnitude less than the threshold value at the centre of the colour bar ($\pm 1$ W m$^{-2}$). For locations where the magnitude is greater than this threshold value, stippling indicates differences that are statistically significant at a significance level of 0.05 after controlling the false discovery rate (Benjamini & Hochberg, 1995; Wilks, 2016); the two-tailed *p*-values are generated by a two-sample *t*-test, assuming equal population variances, using annual mean data from each simulation year as the input; the approximate *p*-value threshold ($p_{FDR}$), which takes the false discovery rate into account, is written beneath the map. The global area-weighted mean difference is also written beneath the map.

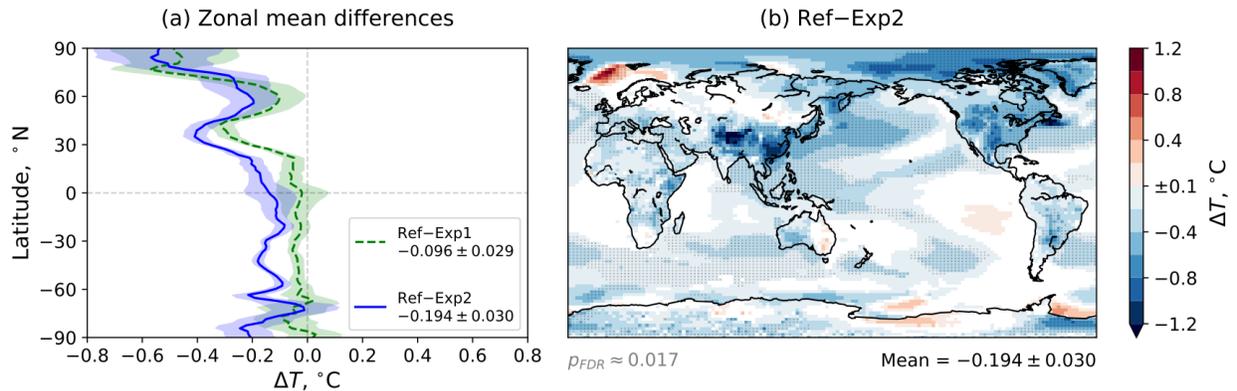

**Figure 3.** Differences in annual mean surface temperature (*T*) for the coupled atmosphere–ocean simulations. The figure components are explained in the caption of Fig. 2.

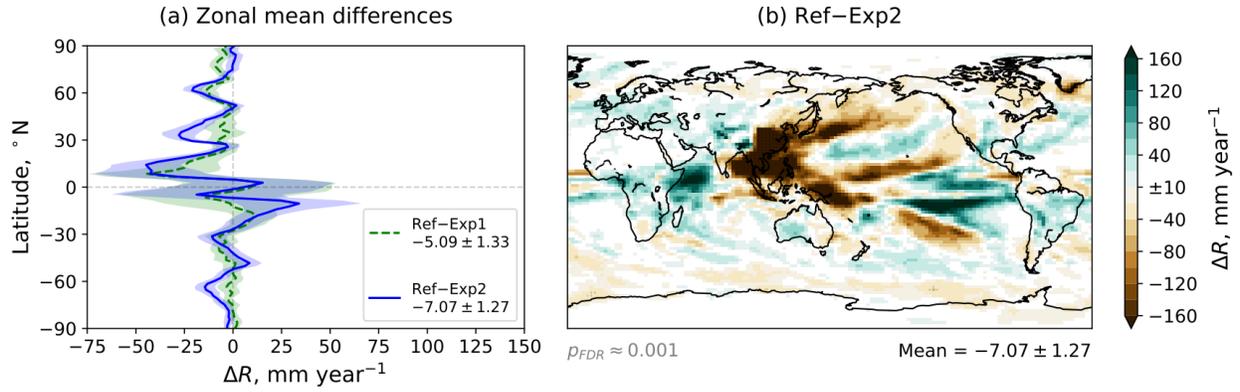

**Figure 4.** Differences in annual total precipitation rate (*R*) for the coupled atmosphere–ocean simulations. The figure components are explained in the caption of Fig. 2.

Supporting Information for

# The Equilibrium Climate Response to Sulfur Dioxide and Carbonaceous Aerosol Emissions from East and Southeast Asia

Benjamin S. Grandey[1], Lik Khian Yeo[1,2], Hsiang-He Lee[1], and Chien Wang[3,1]

[1]Center for Environmental Sensing and Modeling, Singapore-MIT Alliance for Research and Technology, Singapore, Singapore.

[2]Department of Civil and Environmental Engineering, National University of Singapore, Singapore, Singapore.

[3]Center for Global Change Science, Massachusetts Institute of Technology, Cambridge, Massachusetts, USA.

**Contents of this file**

    Introduction
    Figures S1 to S12
    Table S1
    References

**Introduction**

This file contains supplementary figures and a table. These figures and table are referenced in the main manuscript. The scenarios and model configuration are described in the main manuscript. The analysis code is available via https://github.com/grandey/p17d-sulphur-eas-eqm, archived at https://doi.org/10.5281/zenodo.1211416.

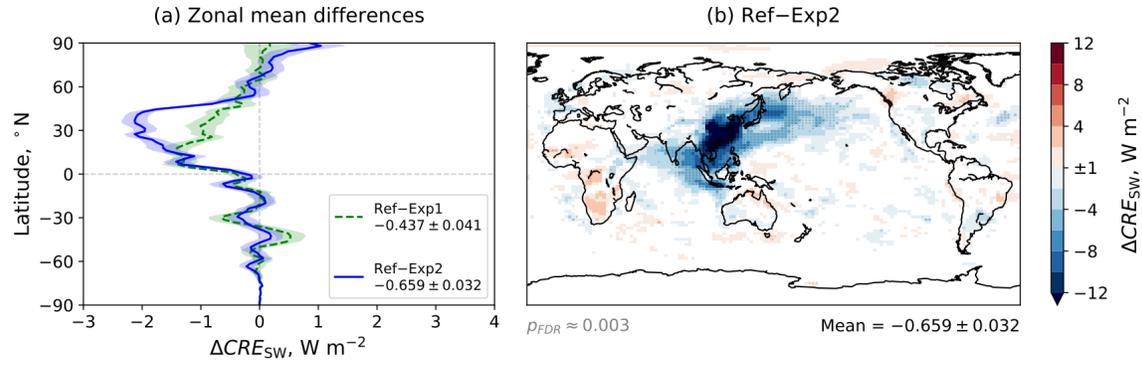

**Figure S1.** Differences in annual mean clean-sky shortwave cloud radiative effect ($CRE_{SW}$), diagnosed using the prescribed-SST simulations. The figure components are explained in the caption of Fig. 2 in the main manuscript.

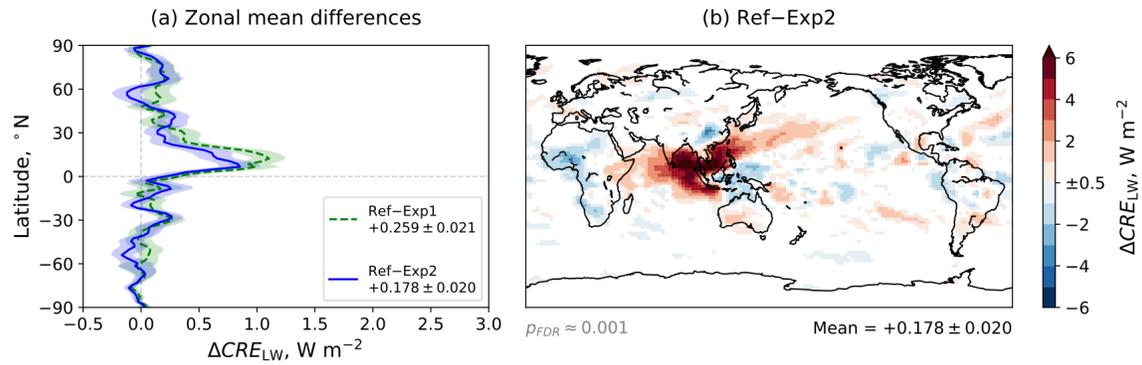

**Figure S2.** Differences in annual mean longwave cloud radiative effect ($CRE_{LW}$), diagnosed using the prescribed-SST simulations. The figure components are explained in the caption of Fig. 2 in the main manuscript.
2

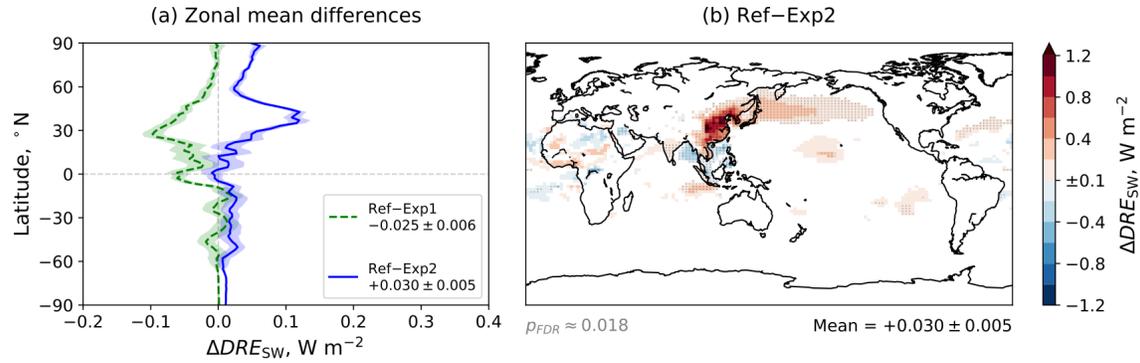

**Figure S3.** Differences in annual mean direct radiative effect ($DRE_{SW}$), diagnosed using the prescribed-SST simulations. The figure components are explained in the caption of Fig. 2 in the main manuscript.

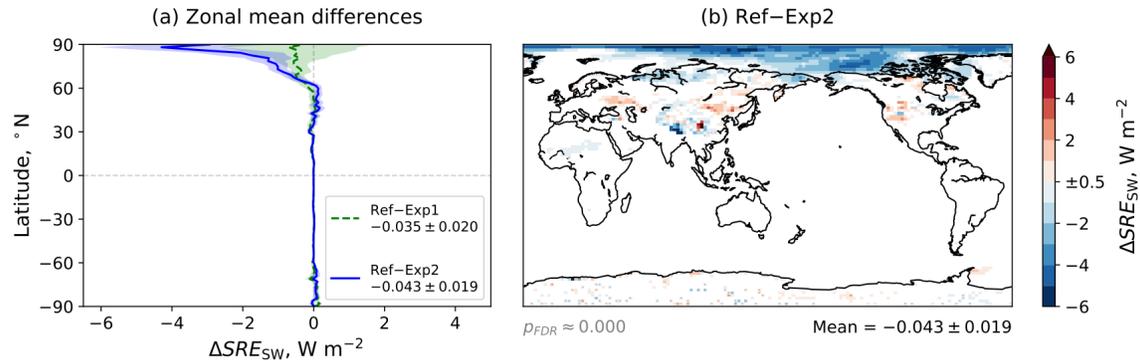

**Figure S4.** Differences in annual mean surface albedo radiative effect ($SRE_{SW}$), diagnosed using the prescribed-SST simulations. The figure components are explained in the caption of Fig. 2 in the main manuscript.

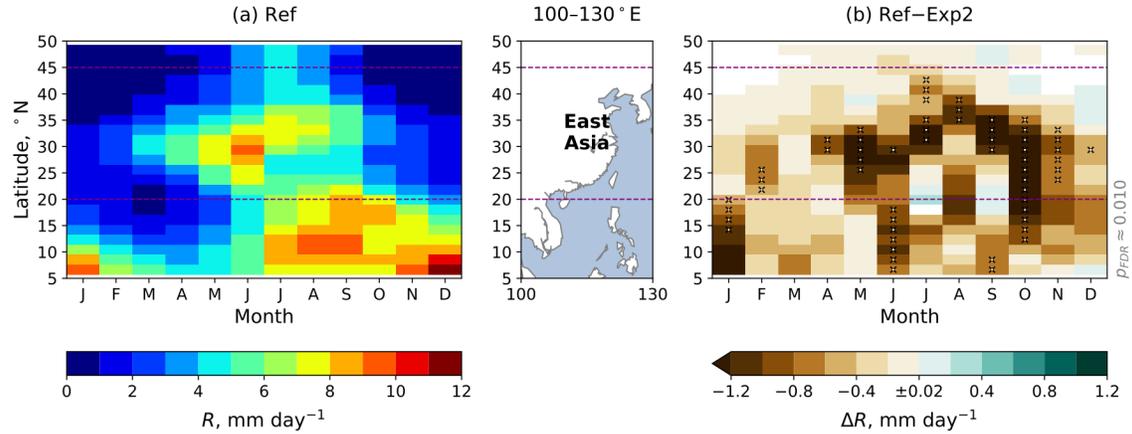

**Figure S5.** Seasonal cycle of total precipitation rate (*R*), zonally averaged across 100–130°E: these longitude bounds, together with the latitude bounds indicated by the purple dashed lines, correspond to the East Asia region of Grandey et al. (2016). (a) Hovmöller diagram for the *Ref* scenario; (b) Hovmöller diagram for *Exp2-Ref* differences. White indicates differences with a magnitude smaller than the threshold value at the centre of the colour bar (±0.2 W m$^{-2}$). For months and latitudes where the magnitude is greater than this threshold value, crosses indicate differences that are statistically significant at a significance level of 0.05 after controlling the false discovery rate (Benjamini & Hochberg, 1995; Wilks, 2016); the two-tailed *p* values are generated by a two sample *t*-test, assuming equal population variances, using zonal mean data from each simulation year as the input; the approximate *p* value threshold ($p_{FDR}$), which takes the false discovery rate into account, is written to the right of the Hovmöller diagram. Both land and ocean data are included in the zonal averages.

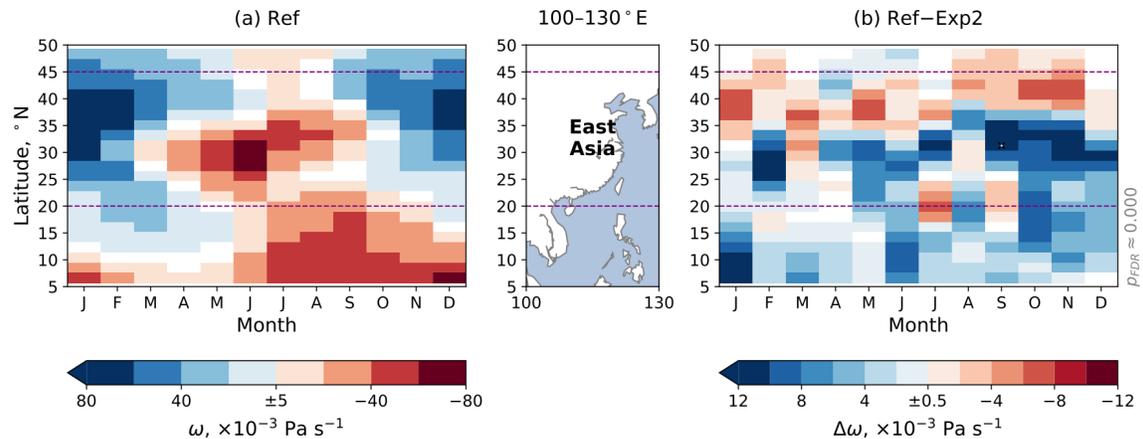

**Figure S6.** Seasonal cycle of vertical motion (*ω*) at model level 19 (~525 hPa), zonally averaged across 100–130°E. Positive values of *ω* indicate downward motion. The figure components are explained in the caption of Fig. S5.

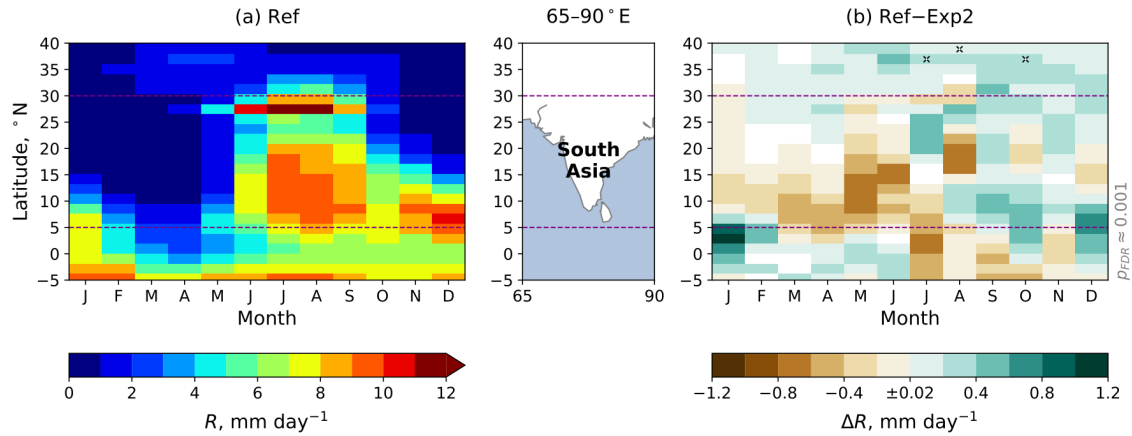

**Figure S7.** Seasonal cycle of total precipitation rate ($R$), zonally averaged across 65–90°E: these longitude bounds, together with the latitude bounds indicated by the purple dashed lines, correspond to the South Asia region of Grandey et al. (2016). The figure components are explained in the caption of Fig. S5.

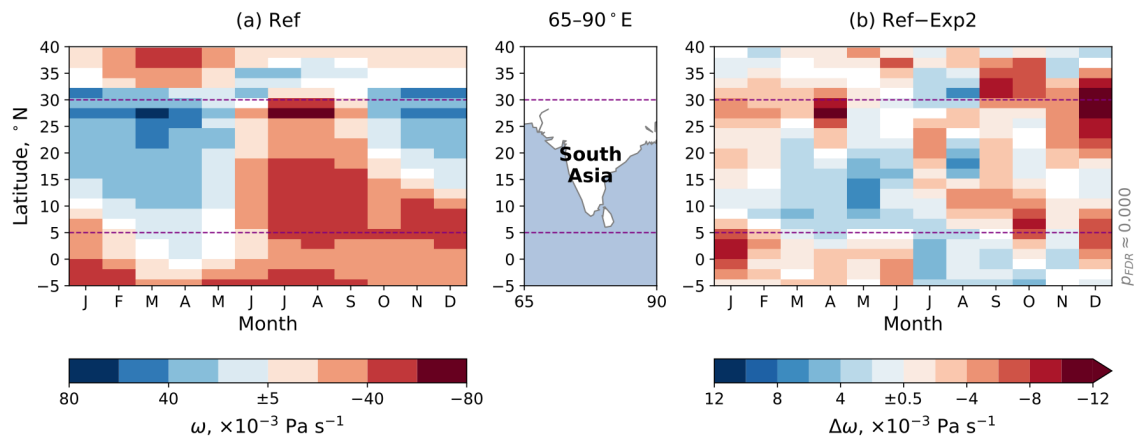

**Figure S8.** Seasonal cycle of vertical motion ($\omega$) at model level 19 (~525 hPa), zonally averaged across 65–90°E. The figure components are explained in the caption of Fig. S5.

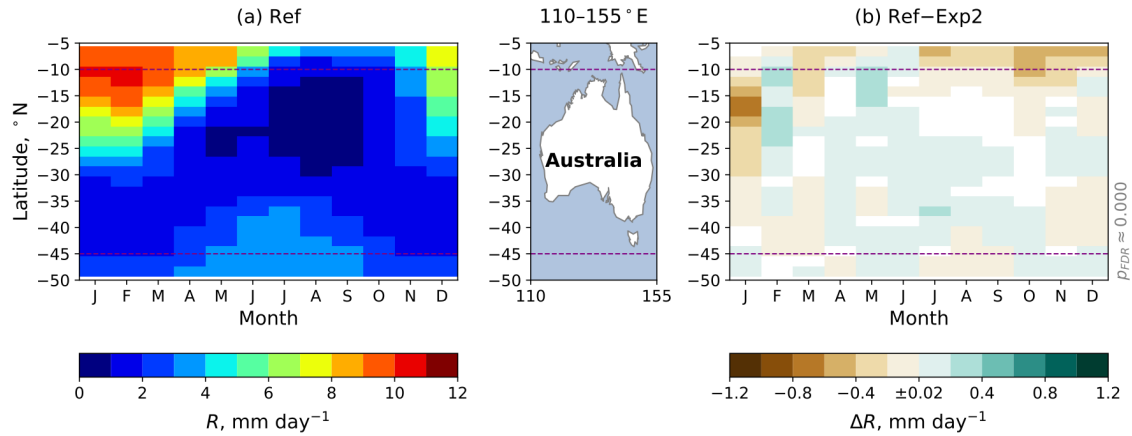

**Figure S9.** Seasonal cycle of total precipitation rate (*R*), zonally averaged across 110–155°E: these longitude bounds, together with the latitude bounds indicated by the purple dashed lines, correspond to the Australia region of Grandey et al. (2016). The figure components are explained in the caption of Fig. S5.

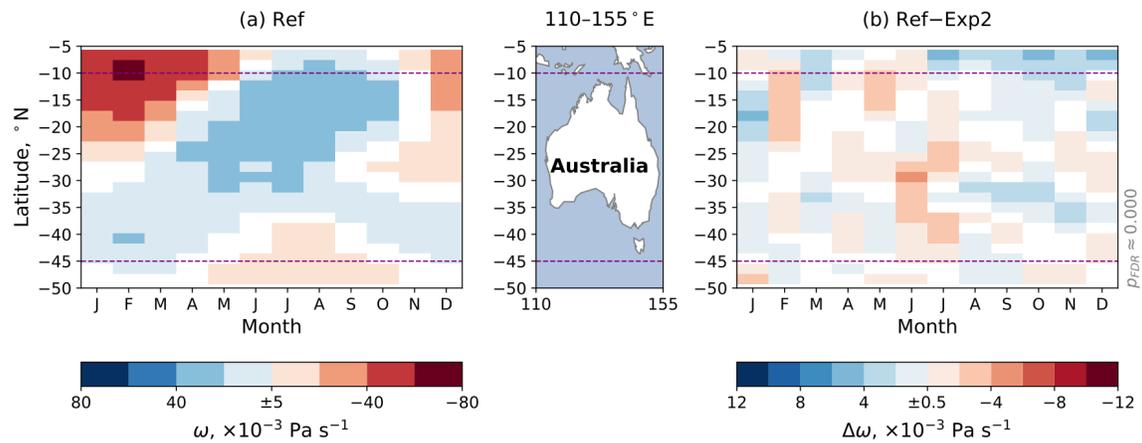

**Figure S10.** Seasonal cycle of vertical motion ($\omega$) at model level 19 (~525 hPa), zonally averaged across 110–155°E. The figure components are explained in the caption of Fig. S5.

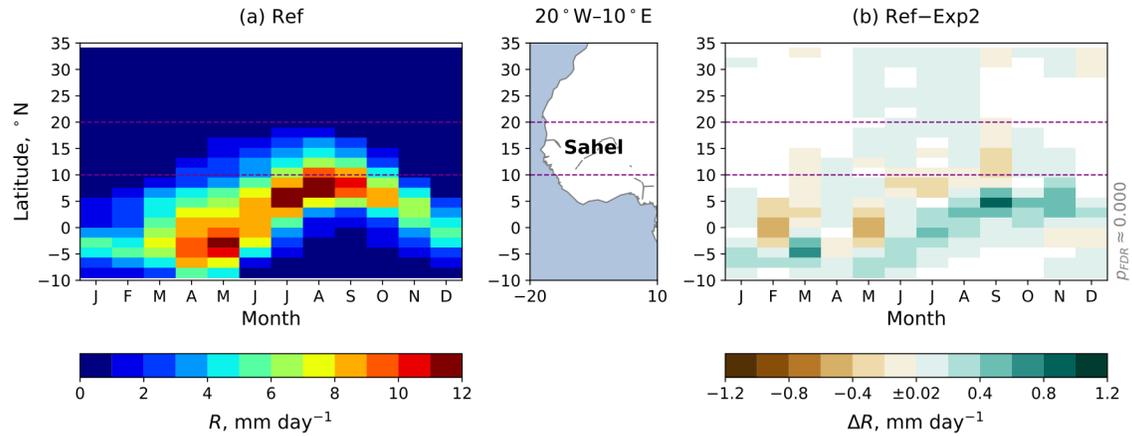

**Figure S11.** Seasonal cycle of total precipitation rate ($R$), zonally averaged across 20°W–10°E: these longitude bounds, together with the latitude bounds indicated by the purple dashed lines, correspond to the Sahel region of Grandey et al. (2016). The figure components are explained in the caption of Fig. S5.

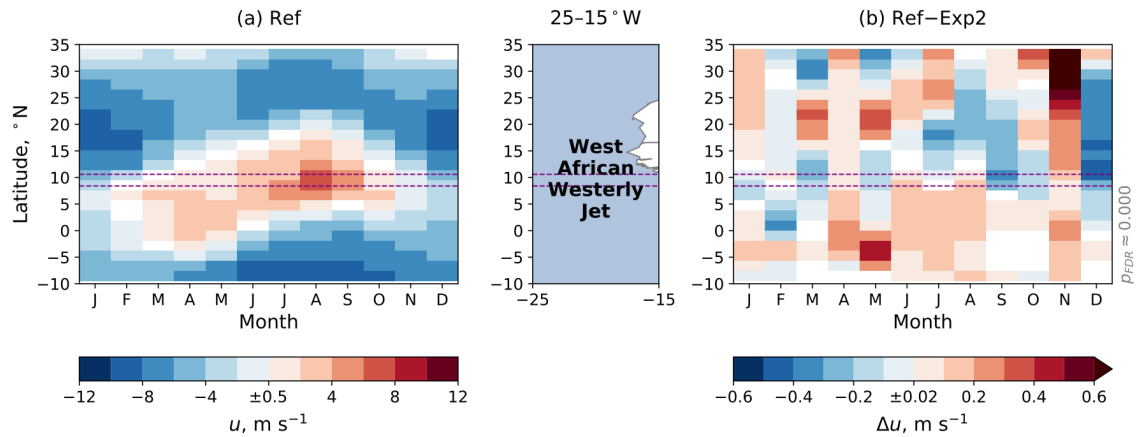

**Figure S12.** Seasonal cycle of zonal wind ($u$) at model level 27 (~936 hPa), zonally averaged across 25–15°W: these longitude bounds, together with the latitude bounds (8.4–10.6°N) indicated by the purple dashed lines, correspond to the region over which 925 hPa zonal wind is averaged in order to calculate the strength of the West African westerly jet (Pu & Cook, 2012). The figure components are explained in the caption of Fig. S5.
7

|  | Grandey et al. (2016) | Current study |
|---|---|---|
| Model version | CESM 1.0.4 with CAM5.1 | CESM 1.2.2 with CAM5.3 |
| Modified emissions region | "Asia": 60–160°E, 10°S–50°N | "East and Southeast Asia": 94–161°E, 10°S–65°N |
| Aerosol emissions scenarios | Two transient scenarios for the 21st Century: (1) RCP4.5 – aerosol emissions projected to decrease globally; (2) A2x – aerosol (sulfur dioxide, organic carbon, black carbon) emissions from Asia projected to increase | Three equilibrium scenarios: (1) Ref – year-2000 emissions; (2) Exp1 – reduced sulfur dioxide emissions; (3) Exp2 – reduced aerosol (sulfur dioxide, organic carbon, black carbon) emissions |
| Greenhouse gas concentrations | Transient, following RCP4.5 | Equilibrium, following year-2000 |
| Significance testing for differences shown on maps and Hovmöller diagrams | Two-sample $t$-test with a two-tailed $p$ value threshold of 0.05 | Two-sample $t$-test with a more rigorous $p$ value threshold, taking the false discovery rate into account |

**Table S1.** Methodological differences between the study of Grandey et al. (2016) and the current study.